\documentclass[a4paper, 10pt]{article}
\usepackage{ctex}
\usepackage{amssymb}
\usepackage{amsmath}
\usepackage{graphics}
\usepackage{epsfig}
\usepackage{subfigure}
\usepackage{cite}
\usepackage[left=2.5cm, right=2.5cm]{geometry}
\usepackage[font={footnotesize}]{caption}

\begin{document}

\def\figurename{{\bf{Fig.}}}

\title{An analytic method for sensitivity analysis of complex systems}

\author{Yueying Zhu$^{1,2,*}$, Qiuping Alexandre Wang$^{1,3}$, Wei Li$^{2,4}$ and Xu Cai$^{2}$ \vspace{0.2cm}\\
\small $^1$ IMMM, UMR CNRS 6283, Universit\'e du Maine, 72085 Le Mans, France\\
\small $^2$ Complexity Science Center \& Institute of Particle Physics,\\
\small Central China Normal University, 430079 Wuhan, China\\
\small $^3$ HEI-ISA-ISEN Group of Engineering Schools, 59046 Lille, France\\
\small $^4$ Max-Planck Institute for Mathematics in the Sciences,\\
\small Inselst. 22, 04103 Leipzig, Germany\\
\small $^*$ Correspondence author: Yueying.Zhu.Etu@univ-lemans.fr
}

\date{}

\maketitle

\abstract{
Sensitivity analysis is concerned with understanding how the model output depends on uncertainties (variances) in inputs and then identifies which inputs are important in contributing to the prediction imprecision. Uncertainty determination in output is the most crucial step in sensitivity analysis. In the present paper, an analytic expression, which can exactly evaluate the uncertainty in output as a function of the output's derivatives and inputs' central moments, is firstly deduced for general multivariate models with given relationship between output and inputs in terms of Taylor series expansion. A $\gamma$-order relative uncertainty for output, denoted by $\mathrm{R^{\gamma}_v}$, is introduced to quantify the contributions of input uncertainty of different orders. On this basis, it is shown that the widely used approximation considering the first order contribution from the variance of input variable can satisfactorily express the output uncertainty only when the input variance is very small or the input-output function is almost linear. Two applications of the analytic formula are performed to the power grid and economic systems where the sensitivity of both actual power output and Economic Order Quantity models are analyzed. The importance of each input variable in response to the model outputs is quantified by the analytic formula.
}



\section{Introduction}\label{sec:1}
Consider a deterministic model $y=f({\bf X})$ with ${\bf X}$ indicating a multivariate vector. When $y$ is calculated from ${\bf X}$ through a specified function, uncertainties in the input variables will propagate through the calculation to the output $y$ \cite{May1976,Huang2015}. This process is called variance propagation (or uncertainty propagation). Variance propagation, which is regarded as the basis of sensitivity analysis for complex models, mainly considers the determination of output's variance via uncertainties in input variables \cite{Blower1994, Gulyuz2016}.

Many methods have been proposed for variance propagation, such as simulation-based methods\cite{Melchers1989,Kiureghian1996}, most probable point-based methods\cite{Hasofer1974,Fiessler1979}, functional expansion-based methods\cite{Xiu2003}, numerical integration-based methods\cite{Seo2002,Xu2004,Rahman2004,Lee2006}. Simulation-based methods, also called sampling-based methods, are regarded as both effective and widely used, especially for those models without specific correspondence between $y$ and ${\bf X}$ \cite{Blower2000,Heltona2006,Marcot2015}. These methods, however, are computationally expensive, especially in the presence of a high number of input variables. For a general model with specific functional relationship between $y$ and ${\bf X}$, the process will be much easier and numerically cheaper for determining the output's variance if an analytic formula associated with variance propagation can be provided. More information associated with other methods for variance propagation can be found in the reviewed papers\cite{Padulo2007,Lee2009,Borgonovo2016}.

A simple analytic formula has been appeared since 1953 which approximately computes the variance of the product of two independent random variables \cite{Yates1960}. In 1966, this approximation was extended by engineers and experimentalists to more general multivariate cases \cite{Ku1966}. This formula, also called Taylor series approximation, restricted to first-order terms \cite{Helton1993}, has gained a wide applications thanks to its simplicity and convenience \cite{White2004}. However, it can satisfactorily estimate the output's variance only when the functional relationship between output and input variables is almost linear or the variance of each input variable is very small \cite{Padulo2007}. For most models, however, $y$ highly nonlinearly depends on ${\bf X}$ having large uncertainties. This suggests the necessity of proposing an analytic formula to exactly calculate the output's variance and then to study its sensitivities in response to different input variables.

In the present paper, an analytic formula for variance propagation is proposed based on Taylor series expansion (univariate case is firstly considered) allowing to exactly determine the output uncertainty as well as the contributions of different orders from input uncertainty in terms of the output's derivatives and input's central moments. This formula is then extended to the general multivariate case followed by the applications in sensitivity and reliability analyses of two complex systems.

The paper is organized as follows: section \ref{sec:2} shows the derivation of the analytic formula for variance propagation and its implementation in different univariate nonlinear functions. The analytic formula is extended to the general multivariate situation in section \ref{sec:4}, with applications in the analyses of two complex systems accompanied. Section \ref{sec:5} concludes the results.

\section{Analytic expression for variance propagation}\label{sec:2}
Beginning with the univariate function, namely $y=f(x)$, its Taylor series expansion about a point $x=\mu$ is provided by
\begin{equation}
\label{taylor:one1}
y=f(\mu)+\sum_{i=1}^{n}\frac{1}{i!}(\frac{d^i f}{d x^i})(x-\mu)^i,
\end{equation}
in which $\mu$ indicates the mathematical expectation of $x$. The above equation holds for a general function connecting $y$ and $x$ by making $n$ go to infinity. Taking the average of both sides of Eq.(\ref{taylor:one1}) yields
\begin{equation}
E(y)=f(\mu)+\sum_{i=1}^{n}\frac{1}{i!}(\frac{d^i f}{d x^i})\mu_i.
\end{equation}
$\mu_i$ is the $i^{th}$ central moment of variable $x$ with definition given by
\begin{equation}
\label{defmk}
\mu_i=\int(x-\mu)^iP(x)\, \text{d}x.
\end{equation}
$P(x)$ labels the probability density function of $x$. The variance of $y$, say $V(y)$, then can be stated as
\begin{equation}
\label{vp:general}
V(y)=\sum_{i,j=0}^{n}\frac{1}{i!\times j!}(\frac{d^i f}{d x^i}\times \frac{d^j f}{d x^j})(\mu_{i+j}-\mu_i\mu_j).
\end{equation}
This formula can not only identify the contributions of different orders of the uncertainty in $x$ with considering different values of $n$, but also exactly determine the output's variance by making $n$ large enough. While $n=1$, Eq. (\ref{vp:general}) only retains the first order contribution from the variance of $x$, indicated as $V(x)$, yielding
\begin{equation}
\label{vp:approximate}
V(y)\approx (\frac{d f}{d x})^2V(x)
\end{equation}
with $\mu_1=0$ and $\mu_2=V(x)$ used. Equation (\ref{vp:approximate}), called the general Taylor series expansion truncated to the first order, is most widely used to approximately calculate the uncertainty in $y$ based on the mean and variance of $x$. This approximation, however, is satisfactory in the frequent case of highly nonlinear functions only when the variance of input is very small\cite{Padulo2007}.

To quantify the contributions of different orders of uncertainty in $x$, a new quantity of interest is proposed, labeled as $\mathrm{R^{\gamma}_v}$, defined as the ratio of $V(y)$ with considering first $\gamma$ orders contributions of uncertainty in $x$ to its exact value:
\begin{equation}
\label{relativevariance}
\mathrm{R^{\gamma}_v}=\frac{V_{\gamma}(y)}{V_{\text{T}}(y)},
\end{equation}
where, $V_{\gamma}(y)$ is calculated from Eq. (\ref{vp:general}) under the condition $i+j\leq 2\gamma$, $V_{\text{T}}(y)$ the theoretical value of $V(y)$ obtained from integral:
\begin{equation}
\label{vp:exact}
V_{\text{T}}(y)=\int(y-E(y))^2P(x)\,\text{d} x.
\end{equation}

Consider a special situation with x following uniform distribution with probability density function given by
\begin{equation}
\label{uniformprobability}
P(x)=
\begin{cases}
\frac{1}{x_m-x_0} \quad &\text{for} \quad x_0\leq x\leq x_m,\\
0 \quad &\text{for}\quad x<x_0\quad \text{or} \quad x>x_m,
\end{cases}
\end{equation}
and $\mu=\frac{1}{x_m+x_0}$. By substituting Eq. (\ref{uniformprobability}) and $\mu$ into Eq. (\ref{defmk}), a generalized representation for central moments is yielded,
\begin{equation}
\label{mk1}
\mu_{2k}=\frac{3^k}{2k+1}V^k(x), \quad M_{2k-1}=0
\end{equation}
with $k$ being a positive integer.

The analytic formula for variance propagation (Eq. (\ref{vp:general})) now can be expressed as a function of output's derivatives and input variance by inserting Eq. (\ref{mk1}):
\begin{equation}
\label{vp:uniform}
V(y)=\sum_{i=1}^{n}\sum_{\mbox{\tiny$\begin{array}{c}
j=i\\
j=j+2\end{array}$}}^{n} C_{ij}
(\frac{d ^iy}{d x^i}\times \frac{d ^jy}{d x^j})V^{(i+j)/2}(x),
\end{equation}
where $j$ is summed with an increment of 2 and $C_{ij}$ is defined as follows:\\
if $i=j$
\begin{equation}
\label{ci}
C_{ij}=\left\{
\begin{array}{cc}
\frac{3^{(i+j)/2}}{(i+j+1)\times i!\times j!} & \text{$i$ and $j$ are odd}\vspace{0.5cm}\\
\frac{3^{(i+j)/2}\times i \times j}{(i+j+1)\times (i+1)! \times (j+1)!} & \text{$i$ and $j$ are even}
\end{array}
\right.
\end{equation}
else
\begin{equation}
\label{cij}
C_{ij}=\left\{
\begin{array}{cc}
\frac{2\times 3^{(i+j)/2}}{(i+j+1)\times i! \times j!}  & \text{$i$ and $j$ are odd}\vspace{0.5cm}\\
\frac{2\times 3^{(i+j)/2}\times i \times j}{(i+j+1)\times (i+1)!\times (j+1)!} & \text{$i$ and $j$ are even}
\end{array}
\right.
\end{equation}

\begin{figure}
\begin{center}
\includegraphics[width=3.5cm]{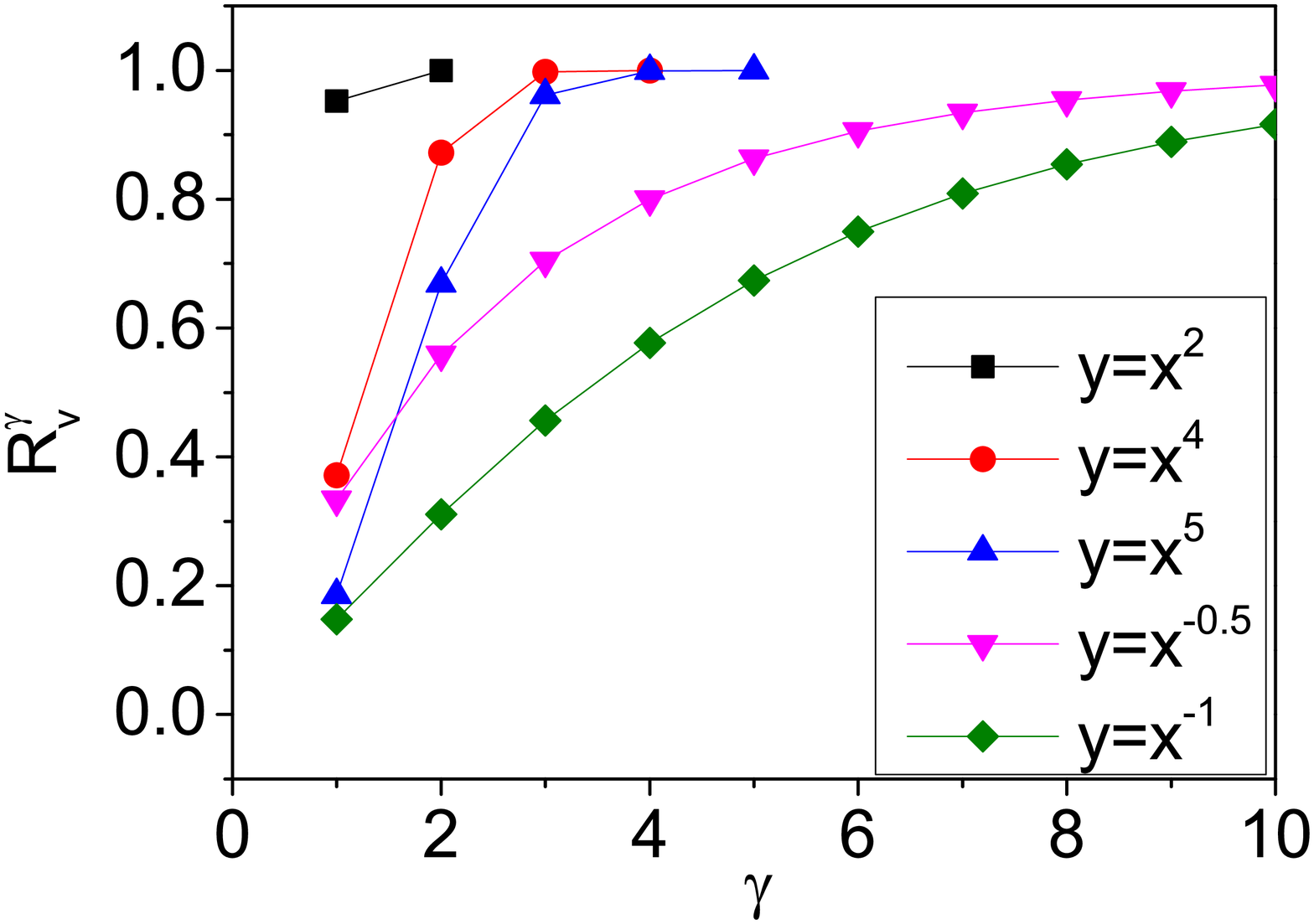}
\includegraphics[width=3.5cm]{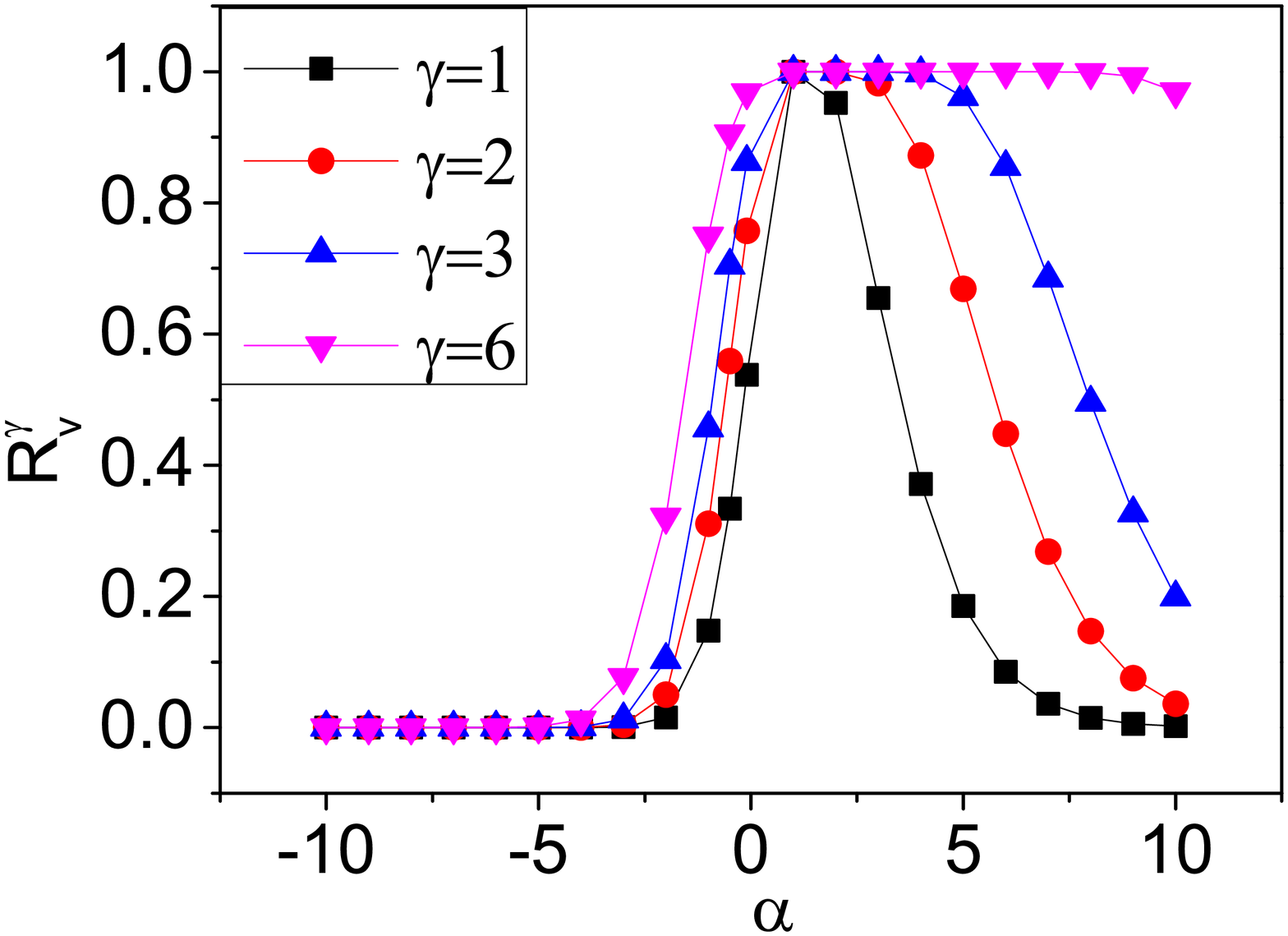}
\includegraphics[width=3.5cm]{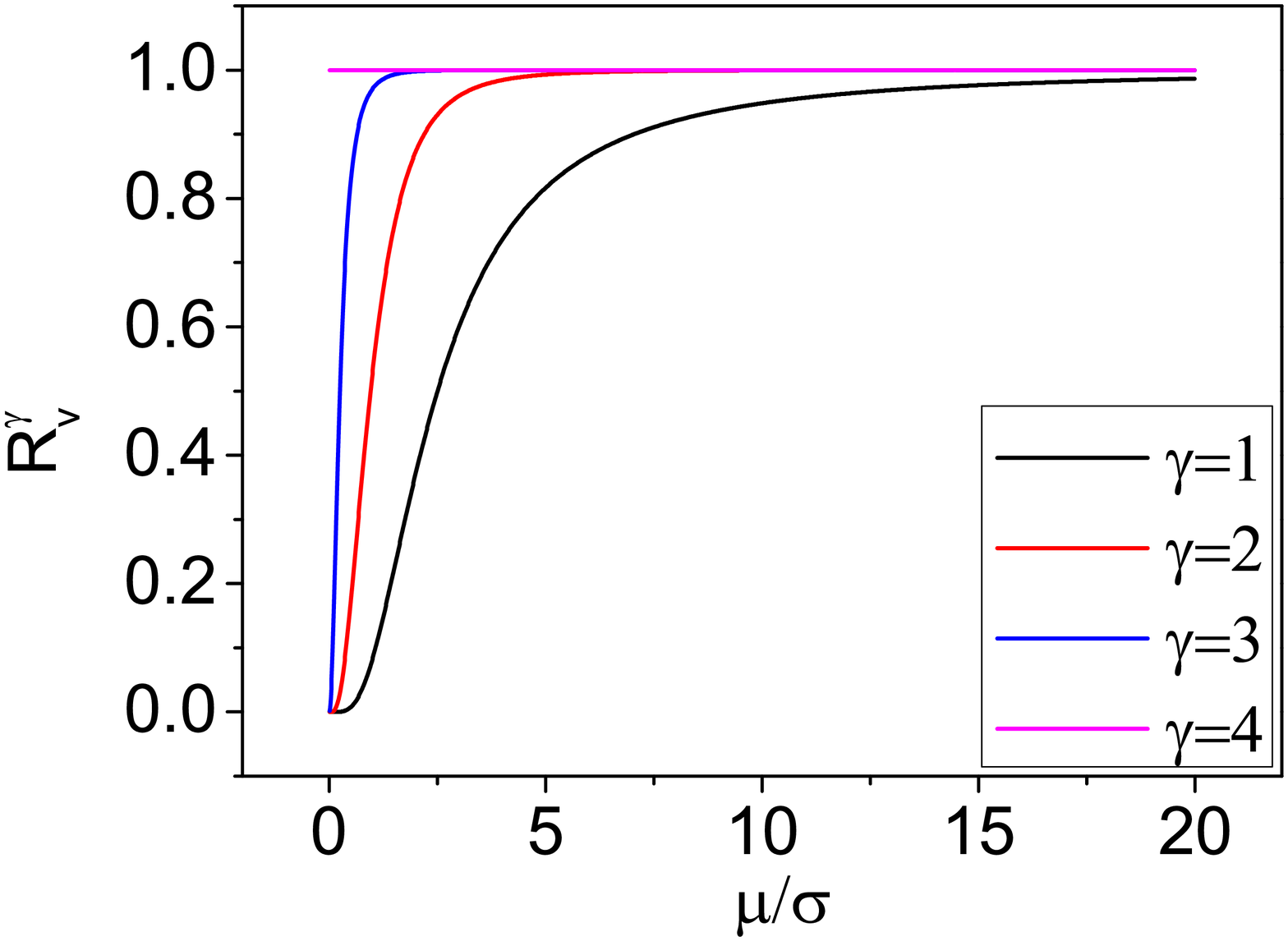}
\end{center}
\centering{$(a)$ Distributions for power-law function}
\begin{center}
\includegraphics[width=3.5cm]{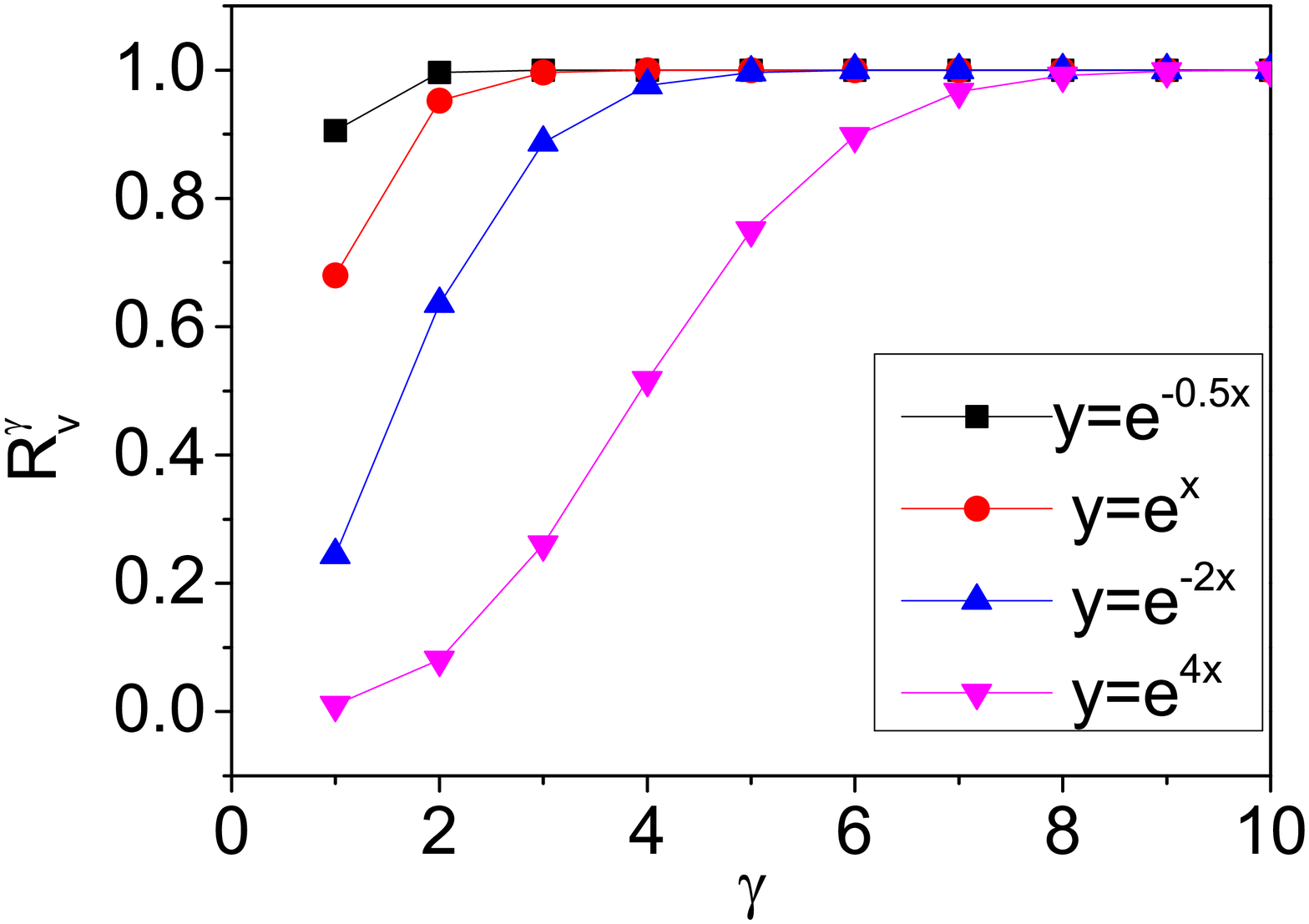}
\includegraphics[width=3.5cm]{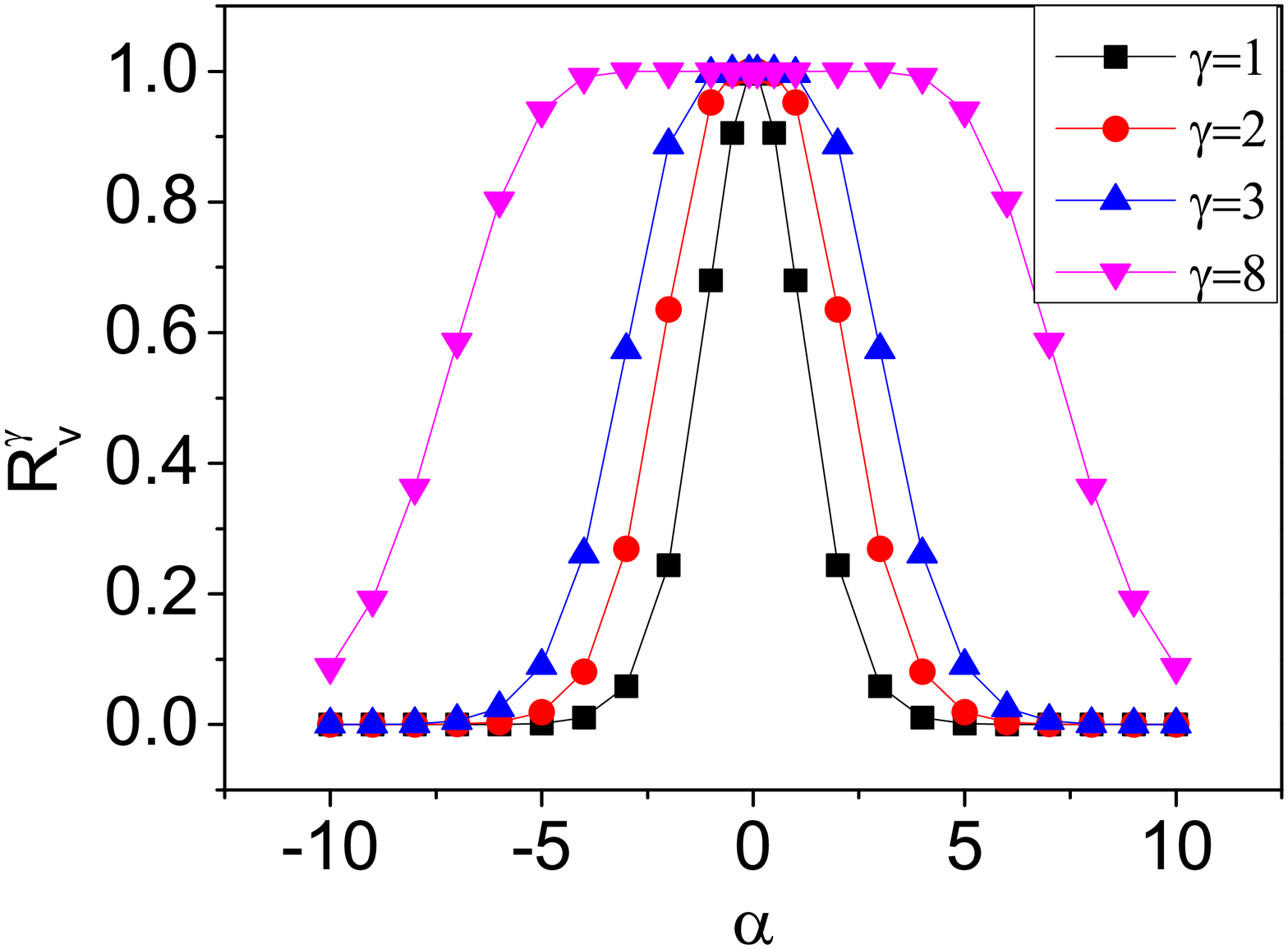}
\includegraphics[width=3.5cm]{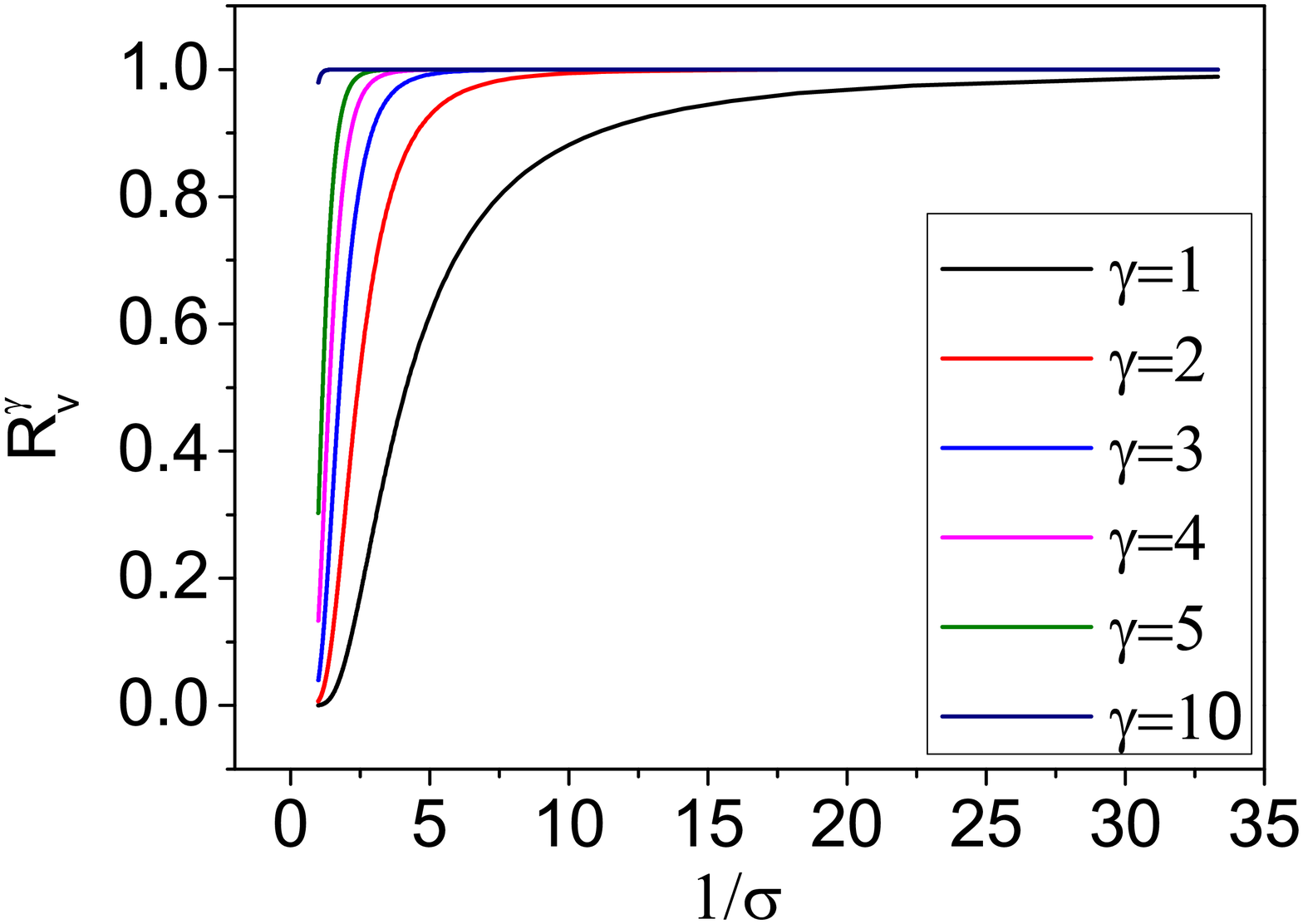}
\end{center}
\centering{$(b)$ Distributions for exponential function}
\caption{The distributions of quantity $\mathrm{R^{\gamma}_v}$ with different parameters for power-law function: $y=x^{\alpha}$ (panel $(a)$) and exponential function: $y=e^{\alpha x}$ (panel $(b)$) with input variable following uniform distribution; To the left in both panels is the distribution of quantity $\mathrm{R^{\gamma}_v}$ with order index $\gamma$ for both classes non-linear functions with different parameter $\alpha$; to the middle is the relationship between $\mathrm{R^{\gamma}_v}$ and parameter $\alpha$ while considering different order contributions from uncertainty in $x$; to the right is the dependence of $\mathrm{R^{\gamma}_v}$ on the distribution parameter of input variable $x$ with $\alpha=4$ for both panels. For a function with specified parameter $\alpha$, $\mathrm{R^{\gamma}_v}$ only depends on the ratio of $\mu$ to $\sigma$ (power-law function) or $\sigma$ (exponential function) for each value of $\gamma$. In the left and middle plots, $\mu/\sigma=2$ in panel $(a)$ and $\sigma^2=0.5$ in panel $(b)$.}
\label{uniform}
\end{figure}

The underlying results of $\mathrm{R^{\gamma}_v}$ for two widely used nonlinear functions are presented in Fig. \ref{uniform}. Apparently, $\mathrm{R^{\gamma}_v}$ can reach 1 while $\gamma$ is large enough for both kinds of functions with different parameters, see left plots in both panels.

By inserting $y=x^{\alpha}$ and $y=e^{\alpha x}$, Eq. (\ref{vp:uniform}) can be respectively updated by
\begin{equation}
\label{term}
\sum_{i=1}^{n}\sum_{\mbox{\tiny$\begin{array}{c}
j=i\\
j=j+2\end{array}$}}^{n}\frac{C_{ij}(\alpha!)^2}{(\alpha-i)!\times (\alpha-j)!}\mu^{2\alpha}(\frac{\sigma}{\mu})^{i+j},
\end{equation}
and
\begin{equation}
\label{term}
\sum_{i=1}^{n}\sum_{\mbox{\tiny$\begin{array}{c}
j=i\\
j=j+2\end{array}$}}^{n}e^{2\alpha \mu} C_{ij}(\alpha \sigma)^{i+j},
\end{equation}
which demonstrate that, for specified $\gamma$,  $\mathrm{R^{\gamma}_v}$ only depends on the ratio of $\mu$ to $\sigma$ (=$\sqrt{V(x)}$, the standard variance of $x$) for power-law function or on $\sigma$ for exponential function while $\alpha$ is fixed. Middle panels in Fig. \ref{uniform} display the distributions of $\mathrm{R^{\gamma}_v}$ with exponent $\alpha$ under specified $\mu/\sigma$ (for power-law function) and fixed $\sigma$ (for exponential function) while considering different order contributions of uncertainty in $x$. Left ones illustrate the dependence of $\mathrm{R^{\gamma}_v}$ upon $\mu/\sigma$ ($1/\sigma$) for power-law function (exponential function) with $\alpha=4$. Regarding to power-law function, $\gamma$ should be larger for larger $|\alpha-1|$ with constant $\mu/\sigma$, or for smaller $\mu/\sigma$ with constant $\alpha$, in order to make $\mathrm{R^{\gamma}_v}$ tend to 1, eg., the considered contributions of uncertainty in $x$ should be until up to the $6^{th}$ order (for making $\mathrm{R^{\gamma}_v>0.98}$) while $\alpha <0.1$ or $\alpha >8$. For exponential function, $\mathrm{R^{\gamma}_v}$ is symmetric with $\alpha=0$ and the contributions of input uncertainty of higher order should be considered when the function more derivatives from linear law or $\sigma$ is larger. The statement is visually verified that the original approximation, Eq. (\ref{vp:approximate}), with just considering the contribution of input uncertainty of first order can successfully estimate the output uncertainty only when the input uncertainty is very small or the considered function is almost linear. Higher order($\gamma \geq 2$) contributions of input uncertainty can not be ignored while regarding highly nonlinear functions when $\sigma$ ($\sigma/\mu$) is large for exponential (power-law) function.

\section{Generalizing the analytic formula}\label{sec:3}

The analytic formula, Eq. (\ref{vp:general}), is only valid for univariate model. However in many mathematical and physical models, the output quantity always depends upon two or more input variables of uncertainty. Hence the generalization of the analytic formula is considered in this section to make it work in the general case with $nX$ independent input variables of the form
\begin{equation}
y=f({\bf X})=f(x_1, x_2, \cdots, x_{nX}),
\end{equation}
which can be similarly expanded by Taylor series:
\begin{eqnarray}
\label{taylor:general}
y&&=f(\{\mu\})+\sum_{t=1}^{nX}\sum_{i=1}^{n}\frac{1}{i!}(\frac{\partial^i f}{\partial x_t^{i}})(x_t-\mu^t)^{i}+\sum_{\substack{t,s=1\\
t<s}}^{nX}\sum_{i_t,i_s=1}^{n}\frac{1}{i_t!\cdot i_s!}(\frac{\partial^{i_t+i_s}f}{\partial x_t^{i_t}\partial x_s^{i_s}})(x_t-\mu^t)^{i_t}(x_s-\mu^s)^{i_s}\nonumber\\
&&+\cdots+\sum_{i_1\cdots i_{nX}}^n\frac{1}{i_1!\cdots i_{nX}!}(\frac{\partial^{i_1+\cdots+i_{nX}}f}{\partial x_1^{i_1}\cdots\partial x_{nX}^{i_{nX}}})(x_1-\mu^1)^{i_1}\cdots (x_{nX}-\mu^{nX})^{i_{nX}}.
\end{eqnarray}
$\{\mu\}$ indicates the mathematical expectation set of input variables: $\{\mu^1, \mu^2, \cdots, \mu^{nX}\}$. Similarly, the variance of output for a general multivariate model then can be exactly calculated by the following expression
\begin{equation}
\label{v:general}
V(y)=\sum_{\substack{i_1\cdots i_{nX}=0\\
j_1\cdots j_{nX}=0}}^{n}\frac{1}{A(i_1,\cdots i_{nX}, j_1,\cdots j_{nX})}\left(\frac{\partial^{i_1+\cdots+i_{nX}}f}{\partial x_1^{i_1}\cdots \partial x_{nX}^{i_nX}}\cdot \frac{\partial^{j_1+\cdots+j_{nX}}f}{\partial x_1^{j_1}\cdots \partial x_{nX}^{j_{nX}}}\right)
\cdot F_{(x_1)^{i_1j_1}\cdots (x_{nX})^{i_{nX}j_{nX}}},
\end{equation}
with $A(\cdots)=i_1!\cdots i_{nX}!\cdot j_1!\cdots j_{nX}!$, and
\begin{equation}
F_{(x_1)^{i_1j_1}\cdots (x_{nX})^{i_{nX}j_{nX}}}=\mu_{i_1+j_1}(x_1)\cdots \mu_{i_{nX}+j_{nX}}(x_{nX})-\mu_{i_1}(x_1)\mu_{j_1}(x_1)\cdots \mu_{i_{nX}}(x_{nX})\mu_{j_{nX}}(x_{nX}).
\end{equation}

\section{Applications in sensitivity analyses of complex systems}\label{sec:4}

The purpose of this section is the applications of the generalized analytic formula, i.e., Eq. (\ref{v:general}), in the sensitivity and reliability evaluations of two complex physical systems: power grid system and economic system. These two systems play extremely important roles in modern societies and their reliability analyses have attracted many researchers' interest.

Regarding to the topic of sensitivity analysis, someone is of most interest to the sensitivity indices. Inspired by the variance decomposition, the output uncertainty can be represented as
\begin{equation}
\label{vd}
V(y)=\sum_{t=1}^{nX}V_{x_t}+\sum_{t=1}^{nX-1}\sum_{s=t+1}^{nX}V_{x_t}V_{x_s}+\sum_{t=1}^{nX-2}\sum_{s=t+1}^{nX-1}
\sum_{u=s+1}^{nX}V_{x_t}V_{x_s}V_{x_u}+\cdots+V_{x_1}V_{x_2}\cdots V_{x_{nX}},
\end{equation}
in which, the first summation set includes the contributions of each input alone, the second one the contributions of the interactions between each two inputs, the third one the contributions of the interactions among each three inputs, and so up to the last one the contribution of the interactions among all $nX$ inputs. All items in Eq. (\ref{vd}) can be computed  by Eq. (\ref{v:general}), eg.,
\begin{align}
\label{vp:sensitivity}
V_{x_t}=&\sum_{i,j=0}^{n}\frac{1}{A(i,j)}(\frac{\partial^iy}{\partial x_t^i}\times \frac{\partial^jy}{\partial x_t^j})\cdot F_{(x_t)^{ij}},\\
V_{x_tx_s}=&\sum_{i,j,k,l=0}^{n}\frac{1}{A(i,j,k,l)}\left(\frac{\partial^{i+k}f}{\partial x_t^i\partial x_s^k}\cdot \frac{\partial^{j+l}f}{\partial x_t^j\partial x_s^l}\right) \cdot F_{(x_t)^{ij}(x_s)^{kl}},\\
V_{x_tx_sx_u}=&\sum_{i,j,k,l,p,q=0}^{n}\frac{1}{A(i,j,k,l,p,q)}\left(\frac{\partial^{i+k+p}f}{\partial x_t^i\partial x_s^k\partial x_u^p}\cdot \frac{\partial^{j+l+q}f}{\partial x_t^j\partial x_s^l\partial x_u^q}\right)\cdot F_{(x_t)^{ij}(x_s)^{kl}(x_u)^{pq}}\\
&-V_{x_t}-V_{x_s}-V_{x_u}-V_{x_tx_s}-V_{x_tx_u}-V_{x_sx_u}.
\end{align}
The sensitivity indices are defined as
\begin{equation}
\label{sensitivitymeasure}
s_t=\frac{V_{x_t}}{V(y)}, \quad s_{ts}=\frac{V_{x_tx_s}}{V(y)}, \quad s_{tsu}=\frac{V_{x_tx_sx_u}}{V(y)}, \quad \cdots,
\end{equation}
which label the sensitivities of output quantity in response to each input alone and to their different order interactions.

\subsection{Power grid system}

Uncertainty and reliability analysis in power grid system has been carried out since 1994 based on Monte Carlo methods \cite{Billinton1994}. The assessment of power grid system reliability is generally divided into two aspects: system adequacy and system security which are respectively related to steady-state operation of system and to the ability of system to withstand sudden natural disturbances or to avoid attack.

In this part, we discuss the reliability of the actual wind power output, namely $P_d$, which is one of the most important items in power grid system, from an analytical view based on our extended formula. $P_d$ depends on two parameters $x$ and $\varepsilon$ through functional relationship \cite{Sansavini2014}:
\begin{equation}
P_d(x)=P(x)+\varepsilon,
\end{equation}
where $x$ labels the wind speed, $P(x)$ the deterministic power output from a wind turbine generator which can be obtained from wind speed:
\begin{equation}
\label{windpower}
P(x)=
\begin{cases}
0 & 0\leq x\leq V_{ci}\\
(A+Bx+cx^2)*P_r & V_{ci}\leq x\leq V_r\\
P_r & V_r\leq x\leq V_{co}\\
0 & x\geq V_{co}
\end{cases}
\end{equation}
and $\varepsilon$ the variation of the power output obeying Gaussian distribution with $\mu=0$ and $\sigma=0.1P_r$ \cite{Jin2010}. Following \cite{Sansavini2014}, we set $V_{ci}=3\text{ms}^{-1}$ and $V_r=12\text{ms}^{-1}$ which respectively denote the cut-in wind speed, at which the turbine first starts to rotate and generate power, and rated wind speed, at which the rated power $P_r$ (the power output limit that the electrical generator is capable of) is reached. The constants $A=0.1215$, $B=-0.0784$ and $C=0.0126$ determined by $V_{ci}$ and $V_r$.

We mainly focus on the reliability evaluation of $P_d$ when $x$ is between $V_{ci}$ and $V_r$. So according to Eq. (\ref{windpower}), the actual power output can be updated as
\begin{equation}
\label{power:system}
P_d(x)=A+Bx+cx^2+\varepsilon.
\end{equation}
In the power grid system, the wind speed $x$ can be represented by the Weibull distribution \cite{Johnson2006}:
\begin{equation}
\rho(x)=\frac{k}{c}(\frac{x}{c})^{k-1}e^{-(\frac{x}{c})^k}; \, (k>0,\, x>0, \, c>1),
\end{equation}
where $c$ and $k$ are the scale parameter and the shape parameter, separately. For simplicity, $c=k=2$ are supposed here. The following quantities then can be yielded
\begin{equation}
\label{Weibulldistribution}
\mu^x=1.97, \, V(x)=5.65, \, \mu_4(x)=150.5.
\end{equation}
By applying the above data, the exact values of power output uncertainty and sensitivity measures now can be obtained:
\begin{equation}
V(P_d)=0.12P_r^2, \, s_x=0.92, \, s_{\varepsilon}=0.08,
\end{equation}
and related analysis results are showed in Table \ref{table1} with the consideration of different $\gamma$ whose maximal value is 2. Apparently, parameter $x$, the wind speed, is much more important to model output $P_d$ with contributing $92\%$ of the uncertainty in $P_d$ compared with parameter $\varepsilon$ who just contributes $8\%$. $s_{x\varepsilon}=0$ indicates no interaction exists between $x$ and $\varepsilon$, which can be understood from Eq. (\ref{power:system}).

\begin{table}
\caption{Analysis results for the power grid system while considering different values of $\gamma$ whose maximal value is 2.}
\label{table1}
\begin{center}
\begin{tabular}{cccccc}
\hline
$\gamma=$ & $V(P_d)$ & $s_x$ & $s_{\varepsilon}$ & $s_{x\varepsilon}$\\
\hline
1 & 0.10$P_r^2$ & 0.90 & 0.10 & 0\\
2 & 0.12$P_r^2$ & 0.92 & 0.08 & 0\\
\hline
\end{tabular}
\end{center}
\end{table}

\subsection{Economic system}

In economic system, one of the oldest classical production scheduling models is the Economic Order Quantity (EOQ) model. This model was developed by Ford W. Harris in 1913 and aims at determining the order quantity that minimizes the total holding costs and ordering costs \cite{Harris1990}. Some analyses about the uncertainty and sensitivity of this model have been proposed in \cite{Schwarz2008,Borgonovo2016}. However, the discussion of its reliability to each input parameter, especially to the interactions between different inputs, is still limited.

This subsection builds an intuition insight into the uncertainty and reliability of EOQ model in terms of the analytic formula deduced before. In EOQ model, the total system cost is expressed as
\begin{equation}
TC=PD+\frac{DK}{Q}+\frac{hQ}{2},
\end{equation}
where $P$, $Q$, $D$, $K$ and $h$ separately denote the purchase unit price, order quantity, annual demand quantity, ordering cost and storage cost. EOQ is the order quantity that minimizes the total system cost. It is easy to obtain the value of $Q$ which determines the minimum point of $TC$:
\begin{equation}
Q^*=\sqrt{\frac{2DK}{h}}.
\end{equation}
The uncertainty of $Q^*$, as well as its sensitivities in response to independent input variables $D$, $K$, $h$ and to their interactions, is quantified in this part by applying Eq. (\ref{v:general}).

Input variables are assumed to be uniformly distributed within the ranges as follows following \cite{Schwarz2008}:
\begin{eqnarray}
&&900\leq D \leq 1600\, \text{unit per year},\nonumber\\
&&\$75\leq K \leq \$125\, \text{per order},\nonumber\\
&&\$1.5\leq h \leq \$2.5\, \text{per order and per year},
\end{eqnarray}
which yields
\begin{equation}
\mu^D=1250,\, \mu^K=100,\, \mu^h=2,\, V(D)=40833.333, \, V(K)=208.333,\, V(h)=0.083.
\end{equation}
Substituting the distribution laws of model inputs into Eq. (\ref{vp:exact}) yields the exact value of output uncertainty: $V(Q^*)=2195$. And the exact values of sensitivity analysis can also be determined:
\begin{equation}
s_D=0.377,\, s_K=0.300,\, s_h=0.314,\, s_{DK}=-0.002, \, s_{Dh}=0.006,\, s_{Kh}=0.005,\, s_{DKh}=0.000.
\end{equation}
Sensitivity analysis results are showed in Table \ref{table2} for different values of $\gamma$. While $\gamma=2$, the analysis results are almost equal to the exact values. This means that the contributions of input uncertainties of $3^{rd}$ or higher-order can be neglected and that the analysis results of $\gamma=2$ can truly represent the reliability of EOQ model. Results show that all three parameters are important to the output. The interaction between each two input parameters also contributes a small part to the uncertainty in output $Q^*$. $s_{DK}=-0.002$ means the interaction between $D$ and $K$ will result in a small decrease of the uncertainty in $Q^*$.

\begin{table}
\caption{Analysis results for the Economic Order Quantity model while considering different values of $\gamma$.}
\label{table2}
\begin{center}
\begin{tabular}{ccccccccc}
\hline
$\gamma=$ & $V(Q^*)$ & $s_D$ & $s_K$ & $s_h$ & $s_{DK}$ & $s_{Dh}$ & $s_{Kh}$ & $s_{DKh}$\\
\hline
1 & 2119	& 0.385	& 0.307	& 0.307	& 0	    & 0	    & 0	    & 0\\
2 & 2192	& 0.377	& 0.300	& 0.314	& -0.002	& 0.006	& 0.005	& 0\\
\hline
\end{tabular}
\end{center}
\end{table}

\section{Conclusions}\label{sec:5}

In the present paper, an analytic formula for variance propagation is proposed. This formula allows to exactly calculate the variance of output variable as a function of the output's derivatives and input's central moments for a general specified univariate function, and can be used for quantifying the contributions of input uncertainty of different orders.

In this work, the formula is applied to two widely used non-linear functions: power-law and exponential functions, considering input variable follows uniform distribution. Results reveal that the widely used approximation with just considering the first order contribution of input uncertainty can satisfactorily express the output variance only for very small input uncertainty or when the input-output relationship is almost linear. For other cases, higher order contributions should be considered for precisely estimating the output variance. This justifies the necessity of an exact formula to quantify different order contributions of input variance to the output uncertainty.

Finally, the proposed formula is generalized to the situation with $nX$ independent input variables. Two applications of the formula are also performed to the power grid and economic systems where the reliability and sensitivity of both actual power output and EOQ model are analyzed. The importance of each input variable to the model outputs is quantified by the analytic formula. This provides some prospectives to analytically identify which parameters are important to the model output for some complex systems. We would like to stress here that this analytic formula is only valid for the case with independent input variables whose probability density functions are specified.

\section*{Acknowledgments}
This work was supported by the Programme of Introducing Talents of Discipline to Universities under Grant No. B08033 and the fellowship from China Scholarship Council under Grant No. 201406770035. The authors would also like to thank Prof. Alain Bulou from Universit\'e du Maine for his constant support and encouragement.


\end{document}